\documentstyle[11pt]{article}
\newcommand{\beq}{\begin{equation}}
\newcommand{\eeq}{\end{equation}}
\newcommand{\beqa}{\begin{eqnarray}}
\newcommand{\eeqa}{\end{eqnarray}}
\newcommand{\bls}[1]{\renewcommand{\baselinestretch}{#1}}
\newcommand{\Title}[1]{\vspace{0mm}
    \begin{center}{\bls{1.3}\Large\bf #1 \\ \bls{1}}\end{center} \vspace{5mm}
}
\newcommand{\Author}[2]{\bls{1.4}
    \begin{center}{{\large {#1}} \\} \bls{1.2} {\small\it #2 \\} \end{center}
}
\newcommand{\Abstract}[1]
{\bls{1.1}
\vspace{3mm}\begin{center}\large\bf Abstract.\end{center}
\par  #1 \vspace*{\fill}
}
\newcommand{\Comment}[1]
{\noindent{\small #1}}

\begin{document}

\Title{
Non-uniform measure in \\ four-dimensional simplicial quantum gravity}

\Author{Bernd Br\"ugmann}
{Physics Department, Syracuse University,\\ Syracuse, NY 13244, USA
\\ bruegmann@suhep.phy.syr.edu}

\Abstract{
Four-dimensional euclidean quantum gravity has been studied as a
discrete model based on dynamical triangulations by Ambj{\o}rn and
Jurkiewicz and by Agishtein and Migdal.  We discuss a particular
implementation of a Monte Carlo simulation of simplicial quantum
gravity. As an application we introduce a non-uniform measure and
examine its effect on simple aspects of the mathematical geometry.
We find that the transition region from the hot to the cold phase is
shifted and that the criticality of the transition changes.
}

\vfill

\Comment{\hfill SU-GP-92/9-1}

\Comment{\hfill hep-lat/9210001}

\bls{1.0}
\newpage
\section{Introduction}

To date there does not exist a single self-consistent theory of
four-dimensional quantum gravity. Even leaving aside profound
conceptual issues like how to understand the concept of time in
quantum cosmology, already at the technical level there are problems
with the quantization of the gravitational field. Perturbation theory,
which has been successfully applied to gauge theories fails for
gravity due to its non-renormalizability. Alternatives to the standard
field theoretic treatment include string theory, but the perturbative
loop expansion is not summable \cite{St}, and it is not clear how to
recover a four dimensional spacetime.  Canonical quantization methods
\cite{DeAs} offer the hope that quantum gravity can be constructed as
a fundamentally non-perturbative theory, but though promising this
approach is far from completion.

Here we will be concerned with an attempt to regularize the formal
euclidean path integral for quantum gravity by a discretization of the
integration over the space of metrics. Such a regularization leads to
a well defined theory, which however may be completely unrelated to
the 'correct' quantum theory of gravity. The relevant question to ask
is therefore whether the so regularized path integral leads to
reasonable physical statements. A much more modest goal is to
understand the statistical model resulting from such a discretization,
and this is what we want to contribute to.

As was shown by Regge \cite{Re} and further explored by others \cite{Ha}, a
classical curved spacetime can be approximated by a piecewise linear
manifold composed of simplices, whose edge lengths define a unique
metric. The traditional Regge calculus fixes the connectivity of the
simplicial complex but allows the edge lengths to vary to obtain all
possible metrics. A related
approach referred to as dynamical triangulations calls for fixing the
edge lengths and varying the connectivity of the simplicial complex
\cite{DyTr}.

The euclidean path integral can be regulated by first of all
fixing the topology of the manifold, replacing the
Einstein action by its simplicial equivalent and by replacing the
infinite dimensional, ill-defined integration over the space of
metrics by a finite summation over simplicial complexes.
This regularization leaves, however, one ambiguity unresolved, which
is the choice of the proper measure in the path integral. The best
way to pick out a unique measure in the continuum theory is through
BRST invariance \cite{Fu}, which results in a path-integral that still
requires a regularization, and the integration variables are no longer
the metric components so triangulations do not seem to be applicable.

It is not known what kind of effective measure the transition to
triangulated spacetimes introduces into the path integral.  In fact,
one reason why the method of dynamical triangulations has become
popular is based on results obtained in two dimensions indicating that
the dynamical triangulated theory is equivalent to exact results
obtained in Liouville theory if in the former the measure factor is
set equal to one \cite{Li,AgBeMiSo}. Naively interpreted, this
suggests that even in four dimensions the summation over dynamical
triangulations somehow takes into account the proper measure factor.
It is therefore of interest to examine whether a non-uniform measure
motivated by the four dimensional continuum theory leads to changes in
the dynamically triangulated theory.

The discretization of the path integral opens the way for a Monte
Carlo simulation. In what follows we will discuss results obtained in
four dimensional simplicial quantum gravity via dynamical
triangulations. For results obtained via Regge calculus see the review
by Hamber \cite{Ha}. There have been two studies in four dimensions,
Ambj{\o}rn and Jurkiewicz \cite{AmJu} and Agishtein and Migdal
\cite{AgMi} (both using the uniform measure). It is still open whether
there is a second order phase transition, which would allow for the
existence of a continuum limit, but in \cite{AgMi}
a diffusion dimension of 4.0 is found in the antigravity phase.

In this article we focus on the relation between the cosmological and
gravitational constants, the net scalar curvature, and the average
distance of two simplices. Since \cite{AmJu} and \cite{AgMi} do not
contain directly comparable data we thought it to be useful to
demonstrate that the computer code written for this article produces
data in agreement with both. The main result of this paper is that the
inclusion of a measure factor of the type $\prod_x
(\mbox{det}g)^{n/2}$ leads to qualitatively different results,
noticeably that the criticality of the transition from the hot to the
cold phase depends on $n$.

In section 2 we introduce the model in detail. In section 3 we comment
on the particular implementation of the data structure representing
the simplicial complex and the Monte Carlo algorithm as far as it
differs from \cite{AmJu,AgMi}. In section 4 we present results
obtained for uniform and non-uniform measures. Section 5
concludes with a discussion.

\section{The Model}

\subsection{Discretized action}

We consider the case where the manifold has the topology of $S^4$. The
euclidean Einstein-Hilbert action is
\beq
   S_E[g] = \lambda \int d^4x \sqrt{g}  - \frac{1}{G} \int d^4x \sqrt{g} R(g),
   \label{ehaction}
\eeq
where $\lambda$ is the cosmological constant, $G$ is the gravitational
constant, $g$ is the determinant of the metric and $R(g)$ its scalar curvature.

As discussed in \cite{AmJu,AgMi}, if one considers only manifolds
which are simplicial complexes with $S^4$ topology and defines the
metric by the condition that all edges have length 1, then the volume
integral $V$ and the net scalar curvature $R$ can be replaced by
\beqa
    V\equiv\int d^4x \sqrt{g} &\longleftrightarrow & N_4[T],
 \label{V} \\
     R\equiv\int d^4x \sqrt{g} R(g) &\longleftrightarrow&
     2\pi N_2[T] - 10 \alpha N_4[T], \label{R}
\eeqa
where $N_i[T]$ denotes the number of $i$-simplices of the
triangulation $T$ and $\alpha$ is derived from the condition that for
approximately flat triangulations the curvature vanishes, $\alpha =$
arccos (1/4) $\approx 1.318$.  The action is then simply \cite{AgMi}
\beqa
    S_E[T] &=& \lambda N_4[T] + \lambda_0  R[T], \\
    \lambda_0 &=& - \frac{1}{G},
\eeqa
setting the relative factor between $V$ and $R$ equal to 1.
Equivalently, the action can be written as \cite{AmJu}
\beqa
    S_E[T] &=& k_4 N_4[T] - k_2 N_2[T],    \label{eha}\\
    k_4 &=& \lambda - 10 \alpha \lambda_0 = \lambda + \frac{10}{G}, \\
    k_2 &=& - 2\pi\lambda_0 = \frac{2\pi}{\alpha G},
\eeqa
where $R$ is replaced by $R/ \alpha$.

Notice that (\ref{eha}) is the most general action of the type $S_E =
\sum_i N_i$ in four dimensions. Euler's relation for $S^4$ and the
Dehn-Sommerville relations leave only two of $N_0, \ldots, N_4$
independent.  The number of vertices $N_0$ in the simplicial complex
for example is
\beq
   N_0 = \frac{1}{2} N_2 - N_4 + 2.
   \label{N0N2N4}
\eeq
In addition, there are inequalities between the $N_i$. Denote by
$o(a)$ the order of vertex $a$, i.e. the number of 4-simplices that
contain $a$. For the average order of simplices we have
\beq
               5  \leq \frac{1}{N_0} \sum_a o(a) = \frac{5N_4}{N_0} < \infty.
\eeq
{}From (\ref{N0N2N4}) and (\ref{R}),
\beqa
    2 < &\displaystyle \frac{N_2}{N_4}& < 4, \\
    -0.614 < &\displaystyle \frac{R}{N_4}& < 12.0, \label{ineqR}
\eeqa
that is, the average curvature is asymmetrically bounded from below
and above.  This is a reflection of the fact that the choice of $S^4$ topology
restricts the space of possible metrics.

The path integral is regulated by
\beq
   Z = \int{\cal D} g e^{-S_E[g]}  \longleftrightarrow
   \sum_{T\in\cal T} e^{-S_E[T]},
   \label{pi}
\eeq
where the integration over all metrics is replaced by a summation over all
triangulations $\cal T$ with $S^4$ topology. Monte Carlo simulations
of the sum over triangulations produce results which are consistent
with a second order phase transition \cite{AmJu,AgMi}, and therefore a
continuum limit may exist. In three dimensions there is a first order
phase transition \cite{ABKV}, while in two dimensions the equivalent of
$S_E$ does not lead to a well-defined theory since only one of the
$N_i$ is independent.

The choice of measure in the path integral is tantamount to defining
the theory. In four-dimensional simplicial quantum gravity no
preferred choice for the measure is known, preferred in the sense that
BRST-invariance singles out a measure in the continuum. Setting the
measure equal to one as in (\ref{pi}), which is used in
\cite{AmJu,AgMi}, is natural only because it is simple, and because in
two-dimensional quantum gravity this uniform measure leads to
numerical results (e.g. \cite{AgBeMiSo}) which are consistent with the
Liouville field theory approach. Of course, there is no reason to
expect that a generalization to higher dimensions should contain the
uniform measure. Rather the opposite is likely, that the
non-triviality of higher dimensions involves the measure, as is
suggested by the increased complexity of higher dimensional
differentiable manifolds.

Since we would like to maintain the metric components $g_{\mu\nu}$ as
integration variables so that a regularization by triangulations is
possible (as opposed to the BRST approach), let us give two examples
for measures in the continuum theory suggested by different physical
arguments. These are
\beqa
    {{\cal D}_1} g_{\mu\nu} &=& \prod_x g^{-5/2} \prod_{\mu\leq\nu}
g_{\mu\nu}, \label{measure1} \\
    {{\cal D}_2} g_{\mu\nu} &=& \prod_x g^{00} g^{-3/2} \prod_{\mu\leq\nu}
g_{\mu\nu}. \label{measure2}
\eeqa
The first one is suggested by diffeomorphism invariance of the measure
\cite{measure} and is scale-invariant. The second is the Leutwyler measure,
which in addition takes into account how the path integral depends on
a particular foliation used in the Hamiltonian formalism, and is also
chosen to cancel certain divergences in the path integral \cite{FrVi}.
Both measures are unsatisfactory because they do not lead to a
well-defined perturbation theory. As an aside, a scale invariant
measure has been considered in the context of quantum Regge calculus
\cite{BeGeMa}, but a direct comparison with our results from dynamical
triangulations does not seem possible.

Here we want to propose that factors of the type $\prod g^{n/2}$ are
of interest even though the true measure is not known and examine the
change they introduce into the Monte Carlo simulations. We make the
assumption that the summation over triangulations does not incorporate
the full measure ${\cal D} g_{\mu\nu} = \prod g^{n/2} g_{\mu\nu}$ but
only the flat part $\prod g_{\mu\nu}$ and hence the equivalent of
$\prod g^{n/2}$ has to be included in the discrete theory.
Notice that as long as the effective measure of summing over all
$T\in\cal T$ is not known --- it probably is neither equivalent to
$\prod g_{\mu\nu}$ nor to $\prod g^{n/2} g_{\mu\nu}$ --- one can at
best hope to get some qualitative insight into the influence of the
measure. Even if the discrete model does depend on the measure,
if there is a second order phase transition, one can attempt
to establish universality of critical exponents under change of
the measure.

The discrete version of the volume element is obtained by
$\sqrt{g}(x)\rightarrow o(a)/5$, which is consistent with (\ref{V})
using $\int d^4x \rightarrow \sum_a$ and $\sum_a o(a) = 5N_4$. The
family of measures including the uniform measure can be incorporated
into the discretized theory by replacing $S_E$ by
\beqa
   S &=& S_E + S_M \\
   S_M &=& - n \sum_a \log \frac{o(a)}{5}.
\label{SM}
\eeqa
One could also imagine different definitions of
local volume, e.g. based on the number of edges sharing a vertex.

\subsection{Monte Carlo simulation}

The Monte Carlo evaluation of the path-integral (\ref{pi}) is largely standard.
One constructs a Markov chain $\{T_i\}$ of triangulations with
equilibrium configuration exp$(-S_E)$ so that
\beq
    \langle f \rangle = \lim_{n\rightarrow\infty} {1\over n}
    \sum_{i=1}^n f(T_i).
\eeq
Of interest is the phase diagram obtained for different values of
$k_2$ and $k_4$. As it turns out, for a given $k_2$ there is a unique
value of $k_4$, $k_4^c(k_2)$, such that if $k_4 < k_4^c$ the volume
$N_4$ goes to infinity and if $k_4 > k_4^c$ then $N_4$ tends to zero.
Hence we fix $N_4$ (canonical simulation) and look for critical points
at some value of $k_2$ on the 'critical' line defined by $k_4 =
k_4^c(k_2)$.

As shown in \cite{GrVa}, the following set of five elementary moves is
ergodic for four dimensional simplicial complexes. Denoting a
4-simplex by its five vertices,
\beqa
&& abcde \longleftrightarrow 1abcd + 1abce + 1abde + 1acde + 1bcde,
\label{m40}\\
&& abcd1 + abcd2 \longleftrightarrow 12abc + 12abd + 12acd + 12bcd,
\label{m31}\\
&& abc12 + abc13 + abc23 \longrightarrow 123ab + 123ac + 123bd,
\label{m2}
\eeqa
where $a,b,\dots$ and $1,2,\ldots$ denote the vertices which are
common to all 4-simplices on the left- and the right-hand side,
respectively. Move $i$ is the exchange of $i$-simplices for
appropriate $(4-i)$-simplices. For example, move 4 as given by going from
left to right in (\ref{m40}) replaces the 4-simplex $abcde$ by five
new simplices obtained by inserting the new vertex 1.

There are two restrictions on when a move may be performed which
ensure that the simplicial complex does not change topology. First of
all, move $i$ is only possible when the order of the simplex which is
to be removed is $5-i$. For move 0, for example, a vertex can only be removed
if it is of order 5, since otherwise its neighboring vertices do not form
a 4-simplex which always has 5 vertices. And second, a move is not
allowed if it creates simplices which are already present. For
example, move 3 --- left to right in (\ref{m31}) --- introduces a new
link 12 that, if already part of the simplicial complex, would lead to
overlapping 4-volumes.

Given this set of ergodic moves, we have to define a random walk
through the space of simplicial complexes and compute the weights
required for detailed balance. Any algorithm to pick a move
introduces a probability $P(T\rightarrow T')$ for a transition from
triangulation $T$ to $T'$. This probability is not independent of the
triangulations involved since the simplicial complex, and therefore for
example the number of allowed moves, is changing.  In general, suppose
the transition probability is
\beq
    W(T\rightarrow T') = P(T\rightarrow T')W_M(T\rightarrow T'),
\eeq
where $W_M$ is the standard Metropolis weight. Detailed balance for
$W$ implies
\beq
    P(T\rightarrow T') = P(T' \rightarrow T),
\eeq
since $W_M$ already satisfies detailed balance.

We choose a prescription for picking a move randomly that does not
require knowledge of the number of possible moves of a given type,
since this number is not readily available in our implementation.
Given a triangulation $T$, first pick a 4-simplex, then any of its
subsimplices with equal probability. The probability of this algorithm
to suggest a particular $i$-simplex for a move of type $i$ from $T$ to
$T'=M_iT$ is
\beq
    P_A(T\rightarrow M_iT) = \frac{o(i\mbox{-simplex})}
    {N_4(T)
    \left(\begin{array}{c}  5 \\ i+1 \end{array} \right)
    },
\eeq
where $o(i\mbox{-simplex})$ is the number of 4-simplices in which the
subsimplex appears. The point is that if a move is allowed, the order
$o$ of the subsimplex must be $5-i$. Since the inverse of move $i$ is
move $4-i$ and
\beq
    f_d(i) := \frac{d-i+1} {
    \left(\begin{array}{c}  d+1 \\ i+1 \end{array} \right)
    } = f_d(d-i),
\eeq
we obtain for the factor $c(T\rightarrow T')$ such that $P = c P_A$
satisfies detailed balance that
\beq
   \frac{ c(T\rightarrow M_iT)} {c(M_iT\rightarrow T)} =
   \frac{N_4(T)}{N_4(M_iT)},
\label{weights}
\eeq
where $N_4(M_iT)=N_4(T)+2(i-2)$. The remaining freedom in the $c(T\rightarrow
M_iT)$ can be adjusted, for example, to speed up thermalization.

The Metropolis weights derived from $S_E$ only depend on the type of
move and not on where it acts on the triangulation, $\Delta S_E(M_i) =
\Delta S_E(i)$ (see \cite{AmJu,AgMi}). To compute the change in $S_M$
the order of the neighboring vertices enters because of the logarithm
in the definition of $S_M$. For a move of type $i$ the change in the
order of old and new common vertices is
\beq
    \Delta o(\mbox{old}) = 2i-5, \;\;\;
    \Delta o(\mbox{new}) = 2i-3.
\eeq
For example, under move 2 (\ref{m2}) the order of vertex $a$, which is
common among the old simplices, changes by -1.

There remain two interesting issues related to ergodicity, one
technical and easily solved, the other fundamental and potentially a
serious problem. First notice that we want to simulate for fixed
values of $N_4$ but that the elementary moves discussed above change
the value of $N_4$, and no set of ergodic moves is known which
preserves $N_4$. The approach we use here is to allow $N_4$ to vary in
a certain range which is a small fraction of $N_4$ but as large as
feasible to approach ergodicity. For details of our implementation see
section 3.

Even though the elementary moves introduced above are ergodic, they
are not finitely ergodic \cite{NaBe}. This is directly related to the
classical result by Markov that there is no algorithm to determine the
type of simplicial 4-manifolds. It is not known whether this leads to
a severe restriction on the space of metrics sampled by the
conventional set of five elementary moves. Of course, for a given
finite number $N_4$ of 4-simplices the elementary moves are finitely
ergodic, but if the infinite volume limit is not just to be
approximated by a finite number of simplicial complexes, then an
algorithm to find all simplicial complexes for all $N_4$ is required and
such an algorithm does not exist.

\section{Computer implementation}

In this section we discuss briefly computer related issues of
Monte Carlo simulations of simplicial quantum gravity using dynamical
triangulations. The reader only interested in the results may skip
this section.

The challenge posed by dynamical triangulations is to find a data
structure for the simplicial complex which allows efficient updating
under the moves (\ref{m40})--(\ref{m2}). As a starting point, suppose
a 4-simplex is represented by the list of its five vertices which are
labeled by integers. A simplicial complex is a list of 4-simplices.

The nature of the five moves requires frequent use of all types of
$i$-simplices, and therefore one is lead to store additional
information about the simplicial complex so that the search for a
particular subsimplex does not become too time consuming. Storing just
vertex labels is one extreme, storing pointers to and maintaining
lists of all $i$-simplices, say, is another. In \cite{AmJu} an
optimized version of the latter approach is employed.

Memory constraints of current workstations may be serious in schemes
which call for storing data about all subsimplices (a 4-simplex has
thirty subsimplices). If 100 bytes are used for each simplex, the
system size is limited to about $N_4=100000$, which is not much
considering that in a flat configuration $N_4/N_0\approx 20$ and we
are dealing with four dimensions. The same holds for
super/parallel computers since the non-linear memory access creates a
worst-case scenario for common caching schemes, so that all the data
should be stored in fast memory.

Another aspect is speed. Clearly, one has to strike a balance between
the speed gained by accessing a subsimplex directly through a pointer
versus the speed lost for updating the data structure.

Notice that the elementary moves all act {\it locally}. Our code
therefore stores for each 4-simplex its five vertex labels and
pointers to its five neighboring simplices. This minimal additional
information allows already to perform each move with only local
searches which are rather fast, while the data structure to be
maintained is small and its updating therefore is also fast and
simple.  Two neighboring simplices share a 3-simplex face of 4
vertices. For the moves the label of the fifth vertex of the
neighboring simplex is needed. This may be one of the reasons why in
\cite{AgMi} for each simplex the five extra vertices of its neighbors
are stored so that one does not have to search for them. The
alternative we use is to store with each simplex the sum of its vertex
labels, and the extra vertices are obtained simply by subtraction.

While the geometric nature of the elementary moves may be complicated,
we have presented them in (\ref{m40})--(\ref{m2}) in a regular form
that suggests a straightforward implementation. Notice that always six
simplices and six vertices are involved. A move is given by a choice
of subsimplex which is to be exchanged. The test whether the order of
the subsimplex is correct can be replaced by the observation that
otherwise one cannot write the suggested move in the regular form, or
equivalently, more than six vertices would be involved. Furthermore,
the simplices appearing in each move are subsets of five out of
possible six vertices in an obvious fashion, and the same is true for
the pointers to neighbors attached to each simplex. Since this
structure is the same for each move, we can use one compact routine that
handles all five moves, which makes debugging simple.

The test for overlapping 4-volumes (the crash test) is performed by a
local recursive search for doubly present subsimplices. Given the new
common subsimplex which will be introduced by a move, all except one
of its vertices are present in the 4-simplex chosen for the move, and
one just has to check whether any of the 4-simplices containing these
vertices also contains the additional vertex of the new common
subsimplex.

To summarize, even though the geometric structure of the dynamic
changes of the triangulation is complicated, there exist a fast,
simple, and memory efficient implementation.

A few comments about the Monte Carlo simulations are in order. The
system is initialized in the minimal $S^4$ configuration, i.e. six
vertices connected in all possible ways, or the six 4-simplices
forming the surface of a 5-simplex. Then moves of type 4 are performed
until a previously fixed value $N_4^0$ is reached. For the reasons
explained in section 2, we perform a constrained simulation such that
$N_4\approx N_4^0$. There are several approaches \cite{Ba,AmJu,AgMi}
that avoid defining a rigid cutoff which might introduce extraneous
effects. They all amount to modifying the action by a potential-like
term with a minimum at $N_4^0$ such that the volume $N_4$ is driven
towards $N_4^0$. We choose
\beq
   S'(k_2,k_4) = \left\{ \begin{array}{ll}
   S(k_2,k_4) - \Delta k_4 N_4 & \mbox{ if $N_4 < N_4^0 - \Delta N_4$}
\\
   S(k_2,k_4) & \mbox{ if $N_4^0-\Delta N_4 \leq N_4 \leq N_4^0 + \Delta
N_4$} \\
   S(k_2,k_4) + \Delta k_4 N_4 & \mbox{ if $N_4 < N_4^0 + \Delta N_4$}
   \end{array}
   \right.
\eeq
where $\Delta N_4$ is the fixed width of the potential well and
$\Delta k_4$ is the fixed slope of the walls of the potential. This
method has also been used with $|x|$ and $x^2$ type potentials.
We checked that the outcome of our simulations does not depend too
sensitively on $\Delta N_4$ and
$\Delta k_4$.

The actual simulations are performed for $k_4^c(k_2)$, for which
the system is in an instable state at $N_4^0$. The modification of $S$
is not introduced as a regulation but serves only to approximate a
canonical simulation with grand canonical elementary moves.

As noted before, if $k_4<k_4^c$, then the geometric constraints are
such that the system moves to larger and larger $N_4$, and vice versa
(which explains the sign in the definition of $S'$). If a good initial
guess $k_4^g$ is available for $k_4^c$, then there are analytic
methods to estimate $k_4^c$ from simulations for $S'(k_2,k_4^g)$
\cite{Ba,AmJu,AgMi}. In our simulations we were confronted with such
widely varying conditions that a fully automated fine-tuning algorithm
for $k_4^c$ was adopted. We use a modified bisection method
to adjust $k_4^g$ until the probabilities for the system to move up or
down near $N_4^0$ are balanced. The modifications take into account
that a high level of noise is present by fixing a minimal bisection
stepsize and allow for the stepsize to increase in case the value of
$k_4^c$ is found to be drifting with thermalization.

\section{Results}

\subsection{Results for uniform measure}

We begin by presenting results for the uniform measure. A typical run
consists of 500 to 5000 sweeps for systems with $N_4^0$ = 4000, 8000
and 16000. A sweep is defined as $N_4^0$ elementary moves not counting
those that are rejected by the Metropolis algorithm or by the
geometric constraints. Error bars were obtained via coarse graining
and not drawn if smaller than the symbol size. We trust the errors for
large $k_2$.  CPU time per run was of the order of 48 hours on a
IBM/RISC 6000.

Figures 1 and 2 show the critical lines $k_4=k_4^c(k_2)$ and $\lambda
= \lambda^c(\lambda_0)$, the latter of which can be directly compared
with \cite{AgMi}, and we find good agreement. The points for $N_4^0$ =
4000 and 8000 are taken at the same set of $k_2$ and often coincide.
In these data there is no indication of a phase transition, the
transition from a crumpled to a smooth phase takes place at a value
away from the minimum in figure 2 (see below).  Thermalization was
very fast, on the order of 20 sweeps.

One may be surprised by the fact that $k_4^c(k_2)$ is almost exactly a
straight line. Notice that since $S_E = k_4 N_4 - k_2 N_2$ the
dominant classical configurations are obtained for $k_4/k_2 = N_2/N_4$,
and $N_2/N_4$ is found to range from 2.0 to 2.5. This is close to the
slopes in figure 1. The constants in the linear transformation between
$k_2$ and $k_4$ and $\lambda_0$ and $\lambda$ happen to be such that
the difference in slope for different $k_2$ becomes much clearer in
figure 2 for $\lambda^c(\lambda_0)$.

Figure 3 and 4 show the curvature per volume, $R/ \alpha V$ from
(\ref{V})--(\ref{R}), and the average geodesic distance $d$ of two
4-simplices, which is the average of the minimal number of steps from
4-simplex to 4-simplex. Both plots are consistent with
\cite{AmJu}. We have noticed that an increase in $\Delta N_4$ (see
section 3) leads to a small shift of the curves to smaller $k_2$,
which might explain why our data is slightly shifted that way in
comparison with \cite{AmJu} where $\Delta N_4 = 0$. Thermalization
ranged from 50 up to 500 sweeps in the transition region.

In figure 3 for $R$, one may suspect that there is a 'kink'
at
\beq
    k_2^c \approx 1.1 \Leftrightarrow \lambda_0 \approx -0.18.
\eeq
The data for $d$ displays clearly a transition from a crumpled phase
for $k_2<k_2^c$ to a smooth phase. When $k_2$ is small, i.e.  $G$ is
large, the average distance does not depend on the size of the system:
essentially any simplex is as close as possible to any other simplex.
If $k_2$ is large, i.e. $G$ is small, then the average distance
increases with the system size: the configuration is smoothed out.
The crumpled and the smooth phase (terminology borrowed from
two-dimensional random surfaces) have also been called hot and cold
phase, respectively.

As observed in \cite{AmJu,AgMi}, the Hausdorff dimension in the
crumpled phase tends to infinity while in the smooth phase it
approaches unity, so the smoothing out happens for extended linear
structures while the crumpling involves higher dimensions, which
explains how more and more simplices can be added without increasing
distances in the system. As an aside, the effective dimension of the
simplicial complex also influences the estimate whether the volume
$N_4$ is small or large for certain purposes. If a system is large
enough in the cold, linear phase that might be far from true in the
hot, crumpled phase.

The curvature per volume is somewhat smaller the larger the
volume. Our data indicates that there are better estimates for the
minimal and maximal curvature per volume than (\ref{ineqR}) (see also
figure 11). As mentioned before, these limits depend on the topology
and are not features of quantum gravity as such. For $S^4$ it is not
clear whether the transition region is far enough removed from the
topology imposed limits so that results can be generalized to
different compact topologies. The average distance of 4-simplices
is larger for larger volumes except for a possible crossover at $k_2 =
1$.

The two phases are also reflected in the acceptance rates of moves,
which are each suggested with roughly the same frequency up to the
factor in (\ref{weights}).  Figure 5 shows the acceptance rates of
moves as the number of moves which had to be suggested for a sweep
when $N_4^0=4000$, and the number of moves accepted by the Metropolis
weight. Typically, even though 70\% of the moves passed the Metropolis
test, only on the order of a few percent satisfy the geometric
constraint. The the test for geometric acceptance should therefore be
highly optimized.

Figure 6 and 7 display for each type of move separately its rate of
geometric acceptance and its percentage of all moves performed due to
the Metropolis weight. Inserting a simplex is always possible (move 4)
and is not shown in figure 6. For example, while the geometry allows
move 3 in about 30\% of all cases, move 2 drops from about 4\% to 1\%.

In figure 7, one recognizes the balance between each move and its
inverse. While in the crumpled phase move 0 is suppressed since the
order of each vertex tends to be large but has to be 5 for the move to
be allowed, in the smooth phase moves 0 and 4 dominate.

\subsection{Results for non-uniform measure}

So far we have restricted ourselves to simple aspects of the
mathematical geometry for uniform measure, which nevertheless clearly
characterize the character of the two phases, and more measurements
can be found in \cite{AmJu,AgMi}. The emphasize will now be on the
influence of a non-uniform measure factor.

In numerical simulations of two-dimensional quantum gravity it was
actually found that the effect of a measure of the type (\ref{SM}) is
negligible since $S_M$ is about constant for fixed measure coupling
\cite{BaJoWi}. Therefore we first measured the value of $s_M=\sum_a
\log o(a)$ in a simulation using the uniform measure. Figure 8 clearly
shows that $s_M(k_2)$ is not constant. At first glance one might
suspect that $s_M(k_2)$ behaves like $N_0$, since a $k_2$-$N_0$ plot
looks exactly like figure 3 for $R$ because of (\ref{N0N2N4}).
However, $R$ may actually become zero and $s_M(k_2)$ must be greater
than a finite positive number, which shows in figure 9 for $s_M/N_0$.

Already from the size of $S_M(k_2)$ for different states of the
geometry one can predict the changes when the non-uniform measure
factor is built into the simulations. Both the very crumpled and very
smooth phase will exist since in these regions $S_M(k_2)$ is constant
because $o(a)$ is constant and the terms proportional to the coupling
constants will dominate $S(k_2)$ assuming that they do not exactly
cancel each other. The contribution from $S_M(k_2)$ is comparable to
that of $S_E(k_2)$ near $k_2=0$. If the factor $n$ is positive, then
$S_M(k_2)$ leads to an effectively larger value of $k_2$ in $S_E(k_2)$
and the transition region will be shifted to smaller values of $k_2$,
and the other way round if $n$ is negative.

Let us see whether the actual simulations agree with these
predictions. Figures 10, 11, and 12 show the $k_2$ dependence of
$k_4^c$, $R/\alpha V$, and $d$ for $n=-5,-1,0,+1,+5$. The curves do
not intersect and show a continuous monotone dependence on $n$. For
$n=-5$ in figure 11 there is evidence for a deformation due to
(\ref{ineqR}). Notice that the curves are qualitatively similar in
scale and shape. This is important because adding a term like $S_M$
could have led to completely new features of the statistical model.
And the curves are indeed shifted as predicted.

The important point is that the shift in origin and the fact that the
curves are stretched in the $k_2$ direction correspond to genuinely
different results. First of all, the supposedly critical point $k_2^c$
ranges from the gravity to the antigravity phase. It is therefore
possible to choose $n$ to obtain particular critical values.

What is potentially much more relevant is that {\it the criticality of the
transition depends on the measure.} If there is a second order phase
transition, then physically relevant statements are obtained for
critical exponents of the transition, and the critical exponents
depend on the criticality of the transition. Figure 11 for $R/\alpha
V$ shows that the slope decreases for both positive and negative large
$n$. The same holds for $d$ in figure 12, where actually the
transition seems to be sharpest for $n= +1$.

The $n$ dependence of $d$ suggests that the Hausdorff dimension
depends on the choice of measure. For fixed $n$ and for large enough
absolute values of $k_2$ the action should become independent of $S_M$
and $d$ should approach its $n=0$ value.

\section{Discussion}

While perhaps not fundamental, simplicial quantum gravity in the
dynamical triangulation approach has the advantage of being an easily
accessible statistical model that may well share some of the features of
full quantum gravity. The question of the effective measure introduced
by the summation over simplicial complexes together with the
ill-understood nature of the measure in the continuum theory render
the ansatz $S=S_E+S_M$ considered in this paper of quite an arbitrary
nature.

However, at least the mathematical geometry can be discussed in some
detail, last but not least because the action $S_E$ is a linear
combination of two global characteristics of the simplicial complex,
say the number of 2- and 4-simplices. As result, the weights in a
Monte Carlo simulation for large $N_4$ depend only on the type of
moves. Thermalization can therefore be discussed for fixed weights but
changing geometry and geometric constraints.  For example, the effect
of the measure term $S_M$ for $n>0$ is to introduce a bias towards
vertices with low order which amounts to increasing the average
distance of 4-simplices and smoothing out the mathematical geometry.

The continuum theory singles out different values of $n$ for different
physical reasons. In the discretized model there also exist preferred
values of $n$, e.g. such that $G$ becomes infinite at the transition
point. Or one may be able to choose $n$ such that the critical value
of $\lambda$ is zero.

The main result is that the criticality of the phase transition
depends on the coupling of the measure term. This we believe may be
quite important for physical predictions based on critical exponents
and is currently under investigation \cite{BrMa}. One clearly has to
study the influence of finite size effects on the measure dependence.

A next step could be to determine the influence of the
measure on some physical definition of the effective dimension of the
simplicial complex (see \cite{AgMi}). If one finds that the dimension
is independent of this type of measure, then a generalized type of
universality may hold. Otherwise the measure could possibly be chosen
to obtain the correct physical dimension.

\vspace{1cm}
{\bf Acknowledgements.} It is a pleasure to thank Enzo Marinari for
many stimulating discussions and the Computational Physics Group and
NPAC of Syracuse for computing resources. This work was supported in
part by NSF grant PHY 90 16733.

\end{document}